\documentclass[twocolumn,amsmath,amssymb]{revtex4-1}
\usepackage{bm}
\usepackage[dvips]{color}
\usepackage{graphicx}
\usepackage{natbib}
\usepackage{url}
\usepackage{amsmath}
\usepackage{float}
\usepackage{xcolor}
\usepackage{epstopdf}
\usepackage{ulem}

\renewcommand{\vec}[1]{\mbox{\boldmath $#1$}}

\begin{document}
\title{Fusion reaction of a weakly-bound nucleus with a deformed target}
\author{Ki-Seok Choi and K. S. Kim}
\address{School of Liberal Arts and Science, Korea Aerospace University, Koyang 412-791, Korea}
\author{Myung-Ki Cheoun}
\address{Department of Physics, Soongsil University, Seoul  156-743, Korea}
\author{W. Y. So}
\address{Department of Radiological Science, Kangwon National University at Dogye, Samcheok 245-905, Korea}
\author{K. Hagino}
\thanks{\textrm{e-mail:} hagino.kouichi.5m@kyoto-u.ac.jp}
\address{Department of Physics, Kyoto University, Kyoto 606-8502, Japan}

\begin{abstract}
We discuss the role of deformation of the target nucleus in the fusion
reaction of the $^{15}$C + $^{232}$Th system at energies around the Coulomb
barrier, for which $^{15}$C is a well-known one-neutron halo nucleus.
To this end, we construct the potential  between $^{15}$C and $^{232}$Th with the
double folding procedure, assuming that the projectile nucleus is composed of the
core nucleus, $^{14}$C, and a valence neutron.
In addition, we also take into account  the coupling to the one-neutron transfer process
to the $^{14}$C+$^{233}$Th configuration.
We show that such calculation simultaneously
reproduces the fusion
cross sections for the $^{14}$C + $^{232}$Th and the $^{15}$C + $^{232}$Th
systems, implying an important role of the transfer coupling in fusion
of neutron-rich nuclei.
\end{abstract}

\maketitle

\section{Introduction}

One of the most important discoveries in nuclei near the neutron drip line
is the halo phenomenon~\cite{Tanihata1985,hansen1987neutron}.
It is characterized by a spatially extended density distribution, originated from
the weakly-bound property of neutron-rich nuclei.
Starting from $^{11}$Li \cite{Tanihata1985},
several other halo nuclei have also been observed successively.
For example, $^{6}$He, $^{14}$Be, and $^{17}$B are regarded as two-neutron halo nuclei,
while $^{11}$Be, $^{15}$C and $^{19}$C are categorized as one-neutron halo nuclei \cite{TANIHATA1985380,PhysRevC.57.2156}.
Recently, heavier halo nuclei, such as $^{19}$C\cite{PhysRevLett.83.1112}, $^{22}$C~\cite{PhysRevLett.104.062701}, $^{31}$Ne~\cite{PhysRevLett.103.262501}, and $^{37}$Mg~\cite{PhysRevLett.112.242501} have also been found at Radioactive Ion Beam
Facility (RIBF) in RIKEN.

Fusion reactions of halo nuclei have attracted lots of attention \cite{TS91,TKS93,HPCD93,DV94,PhysRevC.61.037602,
PhysRevC.65.024606,ITO200653,Canto20061,Canto20151,CHOI2018455,
choi2017estimating}.
It is generally known that fusion cross sections at energies around the
Coulomb barrier are sensitive to the structure of colliding nuclei \cite{BT98,DHRS98,HT12,Back14,MS17}, and it is thus likely that the halo structure
significantly affects fusion reactions, both in a static and a dynamical ways.
With the development of the radio-isotope technology, a large number
of experimental data for fusion of
halo nuclei have been accumulated.
For instance, fusion cross sections for the
$^{11}$Li + $^{208}$Pb~\cite{PhysRevC.87.044603},
$^{\text{6}}$He + $^{238}$U~\cite{raabe2004no},
$^{6}$He + $^{209}$Bi ~\cite{hassan2006investigation,PhysRevLett.81.4580} ,
$^{11}$Be + $^{209}$Bi~\cite{SIGNORINI2004329},
and $^{15}$C + $^{232}$Th~\cite{PhysRevLett.106.172701} systems have
been reported.

Interestingly, it has been reported that
fusion cross sections for the $^{\text{6}}$He + $^{238}$U system~\cite{raabe2004no} do not show any significant influence of the halo structure of
$^{6}$He {\it albeit} that $^{6}$He is a well-known halo nucleus.
This is in contrast to fusion cross sections for the $^{11}$Li + $^{208}$Pb~\cite{PhysRevC.87.044603} system, which show an enhancement
with respect to the fusion cross sections for the $^{9}$Li + $^{208}$Pb system.
The $^{6}$He + $^{209}$Bi system also shows a similar trend
as in the
$^{11}$Li + $^{208}$Pb system ~\cite{PhysRevLett.81.4580}.
In the case one-neutron halo nuclei,
cross sections for the $^{11}$Be+$^{209}$Bi system
are reported to be similar to those for the  $^{10}$Be+$^{209}$Bi system
~\cite{SIGNORINI2004329}.
Origins for this apparent difference among these systems
have not yet been understood completely, even though the fissile nature of
the $^{238}$U may play some role.

In this regard, it is interesting to notice that $^{238}$U is a well deformed
nucleus while $^{208}$Pb and $^{209}$Bi are spherical nuclei.
The aim of this paper is
to investigate the role of deformation of the target nucleus in fusion
of a halo nucleus. To this end,  we shall discuss
the fusion reaction of the $^{15}$C+$^{232}$Th system.
The $^{15}$C nucleus is a one-neutron halo nucleus~\cite{PhysRevC.57.2156},
and its structure is much simpler than the structure of
the two-neutron halo nuclei $^{11}$Li and $^{6}$He.
The $^{15}$C+$^{232}$Th system thus provides an ideal opportunity to
disentangle the deformation and the halo effects.
Moreover, $^{15}$C is heavier than $^{11}$Li and $^{6}$He, and
more significant effects of the target deformation can be expected for
the  $^{15}$C+$^{232}$Th system as compared to the
$^{11}$Li, $^{6}$He+$^{238}$U systems.
Notice that the previous calculation for this system used
a very simple spectator model and did not take into
account  the halo structure of $^{15}$C~\cite{PhysRevLett.106.172701}.
It has yet to be clarified how much the measured fusion enhancement can
be accounted for by taking into account the halo structure of $^{15}$C.

The paper is organized as follows.
In Sec.~\ref{sec2}, we first analyze the fusion of the $^{14}$C + $^{232}$Th system by using a deformed Woods-Saxon potential.
In Sec.~\ref{sec3}, we
analyze the fusion of the $^{15}$C + $^{232}$Th system and
discuss the role of the halo structure and the deformation effect.
To this end, we construct the potential
between $^{15}$C and $^{232}$Th with the double folding formalism
accounting for the halo structure of the $^{15}$C nucleus.
We finally summarize the paper in Sec.~\ref{sec4}

\section{Fusion reaction of the $^{14}$C+$^{232}$T\lowercase{h} system}
\label{sec2}

Before we discuss the fusion cross sections for the  $^{15}$C+$^{232}$Th system,
we first analyze the $^{14}$C+$^{232}$Th system.
In order to take into account the deformation effect
of the target nucleus $^{232}$Th,
we employ a deformed Woods-Saxon (WS) potential for the
relative motion between the target and the projectile nuclei
\cite{HT12,HRK99,PhysRevC.98.014607}:
\begin{equation}
V_{^{14}{\rm C}-T}(r,\theta) = - \frac{V_0}{1+ \exp\left[ \left( r-R_0 - R_T\sum_{\lambda} \beta_{\lambda T} Y_{\lambda 0 }(\theta) \right)/a\right]},
\end{equation}
where $V_{0}$, $R_{0}$, and $a$ are the depth, the radius, and the diffuseness parameters, respectively. $R_T$ and $\beta_{\lambda T}$ are the radius and the
deformation parameters of the target nucleus $^{232}$Th.
The Coulomb potential also has a deformed form given by \cite{HT12,HRK99}
\begin{eqnarray}
V_{C}(r,\theta) &=& \frac{Z_{P}Z_{T}e^{2}}{r} \nonumber\\
& &+\frac{3Z_{P}Z_{T}e^{2}}{5}\frac{R_{T}^{2}}{r^{3}}\left(\beta_{2T}+\frac{2}{7} \sqrt{\frac{5}{\pi}}\beta_{2T}^{2}  \right)Y_{20}(\theta)\nonumber\\
& &+\frac{3Z_{P}Z_{T}e^{2}}{9}\frac{R_{T}^{4}}{r^{5}}\left(\beta_{4T}+\frac{9}{7\sqrt{\pi}}\beta_{2T}^{2}  \right)Y_{40}(\theta),
\label{CoulombV}
\end{eqnarray}
with  the second order in the quadrupole deformation parameter, $\beta_{2T}$,
and the first order in the hexadecapole deformation parameter, $\beta_{4T}$.
$Z_P$ and $Z_T$ are the atomic number of the projectile and the target nuclei,
respectively. Fusion is simulated with the incoming wave boundary condition \cite{HT12,HRK99}.
For simplicity, in this paper we assume that $^{14}$C is inert.

\begin{figure}[h]
\includegraphics[width=0.7\linewidth]{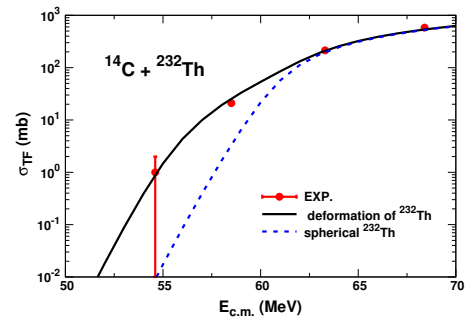}
 \caption{Fusion cross sections for the \(^{14}\)C + \(^{232}\)Th system.
The dashed line denotes cross sections in the absence of the deformation
effect of the target nucleus $^{232}$Th while the solid line is obtained by taking
into account the deformation effect with a deformed Woods-Saxon potential.
The experimental data are taken from Ref.~\cite{PhysRevLett.106.172701}.}
\label{fig:1}
\end{figure}

The fusion cross sections for the $^{14}$C+$^{232}$Th system so obtained
are presented in Fig. \ref{fig:1}.
The actual values for the parameters in the WS potential employed in this
calculation are given in
Table \ref{set-14C}.
We use the same values for the radius and the diffuseness parameters as those
of the global type of Ak\"yz-Winther (AW) potential \cite{akyuz1981proceedings},
while we adjust the depth parameter, $V_0$ to fit the measured fusion cross sections.
For the deformation parameters, we employ
$\beta_{2T} = 0.233$ and $\beta_{4T}= 0.0946$~\cite{zumbro1986}. In the figure,
the blue dashed line shows the cross sections in the absence of the deformation
effect, while the black solid line shows the cross sections
with the deformation effect.
One can clearly see that the enhancement of the fusion cross sections below the Coulomb barrier region
can be well accounted for by taking into account the deformation
of $^{232}$Th.
It is apparent that the deformation plays an important role in this system.

\begin{table}
\begin{ruledtabular}
\caption{The depth, $V_0$, the radius, $r_0$, and the diffuseness, $a$, parameters
for the deformed Woods-Saxon potential for the $^{14}$C + $^{232}$Th reaction.
Here, the radius parameter $r_0$ is defined as $R_0=r_0(A_P^{1/3}+A_T^{1/3}$), where
$A_P$ and $A_T$ are the mass numbers of the projectile and the target nuclei, respectively. The resultant barrier height, $V_{b}$, the barrier position, $R_{b}$, and the barrier curvature, $\hbar \Omega$ are also shown. }
\label{set-14C}
\begin{tabular}{ccc|ccc}
  $V_{0}$ (MeV)   &  $r_{0}$ (fm)  &  $a$ (fm)  &  $V_{b}$ (MeV) & $R_{b}$ (fm)  &  $\hbar \Omega$ (MeV)\\ \hline
     68.174       &    1.231    &   0.548        &   60.66      &  12.22      &     4.96              \\
\end{tabular}
\end{ruledtabular}
\end{table}

\section{Fusion reaction of the $^{15}$C+$^{232}$T\lowercase{h} system}
\label{sec3}

\subsection{The internuclear potential}

Let us now discuss the fusion reaction of the $^{15}$C + $^{232}$Th system.
We first construct the potential between $^{15}$C and $^{232}$Th
taking into account the deformation effect of the target nucleus as well as the
halo structure of the projectile.
To this end, we employ the double folding approach and construct
the potential as
\begin{eqnarray}
V_{^{15}{\rm C}-T}(\vec{r};\vec{r}_{\rm d})
&=& \int d\vec{r}_p \int d\vec{r}_T \rho_p(r_p)\rho_T(\vec{r}_T;\vec{r}_{\rm d})
\nonumber \\
&&\times V_{NN} (\vec{r}-\vec{r}_p+\vec{r}_T),
\label{folding}
\end{eqnarray}
where $V_{NN}$ is an effective nucleon-nucleon interaction, while
$\rho_p(r_p)$ and $\rho_T(\vec{r}_T;\vec{r}_{\rm d})$
are the density profiles for the projectile and the target nuclei, respectively.
Here, the density of the deformed target is defined with respect to the orientation
angle, $\vec{r}_{\rm d}$, in the space fixed frame.
In this paper, we employ the deformed Woods-Saxon density
given by
\begin{equation}
\rho_T(\vec{r};\vec{r}_{\rm d})=
\frac{\rho_0}{1+\exp\left[\left(\frac{r-c(1+\sum_{\lambda} \beta_{\lambda T} Y_{\lambda 0 }(\theta_{rd}))}{z}\right)\right]},
\label{defWSdensity}
\end{equation}
where $\theta_{rd}$ is the angle between $\vec{r}$ and $\vec{r}_{\rm d}$, and
$\beta_{\lambda T}$ are the deformation parameters.
We take $c$=6.851 fm, $z$=0.518 fm and $\rho_{\text{0}}$=0.162 fm$^{-3}$
for the $^{232}$Th nucleus~\cite{DEVRIES1987495}.
For the nucleon-nucleon interaction, $V_{NN}$,
we use the M3Y interaction \cite{BERTSCH1977399}  given by
\begin{equation}
  V_{NN}(r)=-2134\frac{e^{-2.5r}}{2.5r}+7999\frac{e^{-4r}}{4r}-275.81\delta(r),
\end{equation}
where the energy and the length are given in units of MeV and fm, respectively.
Notice that this interaction also includes the knock-on exchange effect in the zero-range approximation.

To evaluate the double folding potential,
the target density (\ref{defWSdensity}) is expanded as
\begin{eqnarray}
\rho_T(\vec{r};\vec{r}_{\rm d})&=&
\sum_{\lambda} \rho_{T \lambda}(r) Y_{\lambda 0}(\theta_{rd}) \\
&=&
\sum_{\lambda,\mu} \rho_{T \lambda}(r) \sqrt{\frac{4\pi}{2\lambda+1}}
Y_{\lambda\mu}(\hat{\vec{r}})Y^*_{\lambda\mu}(\hat{\vec{r}}_{\rm d}).
\end{eqnarray}
Substituting this into Eq. (\ref{folding}), one obtains the potential in a form of
\begin{equation}
V_{^{15}{\rm C}-T}(\vec{r};\vec{r}_{\rm def})
=\sum_{\lambda,\mu} V_{\lambda}(r) \sqrt{\frac{4\pi}{2\lambda+1}}
Y_{\lambda\mu}(\hat{\vec{r}})Y^*_{\lambda\mu}(\hat{\vec{r}}_{\rm d}),
\end{equation}
with
\begin{equation}
 V_{\lambda}(r) = \int d\textbf{r}_p \int d\textbf{r}_T \rho_n(r_p) \rho_{T \lambda} (r_T)
 V_{NN} (\vec{r}-\vec{r}_p+\vec{r}_T).
\label{expansion}
\end{equation}
In the isocentrifugal approximation, one then sets $\hat{\vec{r}}_{\rm d}=0$ and
finally obtains \cite{HT12}
\begin{equation}
V_{^{15}{\rm C}-T}(r,\theta)
=\sum_{\lambda} V_{\lambda}(r) Y_{\lambda 0}(\theta).
\end{equation}
%

We assume that the projectile nucleus $^{15}$C takes the two-body structure,
with the spherical core nucleus $^{14}$C and a valence neutron.
The density of the projectile is then given as
\begin{equation}
\rho_p(\vec{r})=\rho_c(r)+\rho_n(r),
\end{equation}
where $\rho_c(r)$ and $\rho_n(r)$ are the density for the core nucleus and the
valence neutron, respectively.
If one uses this density, the folding potential of Eq.~(\ref{folding}) is
also separated into two parts:
\begin{equation}
V_{^{15}{\rm C}-T}(r,\theta) = V_{^{14}{\rm C}-T}(r,\theta)+V_{n-T}(r,\theta) .
\label{poten-sum}
\end{equation}
For simplicity,
we replace the interaction between the core and the target nuclei,
$V_{^{14}{\rm C}-T}(r,\theta)$, by the deformed Woods-Saxon potential
determined in the previous section.

\begin{figure}[h]
 \includegraphics[width=0.7\linewidth]{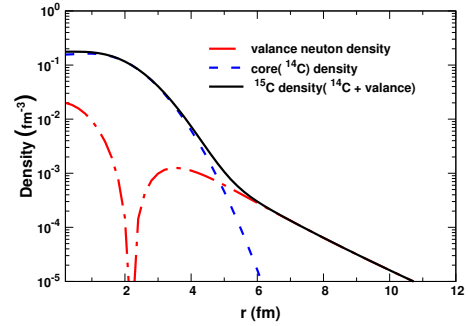}
\caption{The density distribution of the \(^{15}\)C nucleus (the solid line).
The dashed and the dot-dashed lines denote the contribution of the core nucleus
and the valence neutron, respectively. }
\label{fig:2}
\end{figure}

For the density for the valence neutron,
we construct it using a $2s_{1/2}$ neutron wave function in a Woods-Saxon potential
as
\begin{equation}
\rho_n(r)=\frac{1}{4\pi}\,\left[R_{2s_{1/2}}(r)\right]^2,
\end{equation}
where $R_{2s_{1/2}}(r)$ is the radial part of the wave function.
To this end, we use the Woods-Saxon potential with set C in Ref. \cite{HS07}, which
reproduces the empirical neutron separation energy for this state,
$\epsilon_{2s_{1/2}}=-1.21$ MeV.
Figure \ref{fig:2} shows the projectile density thus obtained.
The blue dashed line shows the density for the core nucleus, $^{14}$C, while the
red dot-dashed line denotes the valence neutron density.
For the description of the core density, we use the modified harmonic-oscillator
model, whose parameters can be found in Ref.~\cite{DEVRIES1987495}.
One can see that the valence neutron density has a long tail, reflecting the
halo structure of the $^{15}$C nucleus.


\begin{figure}[h]
\begin{tabular}{ccc}
\includegraphics[width=0.70\linewidth]{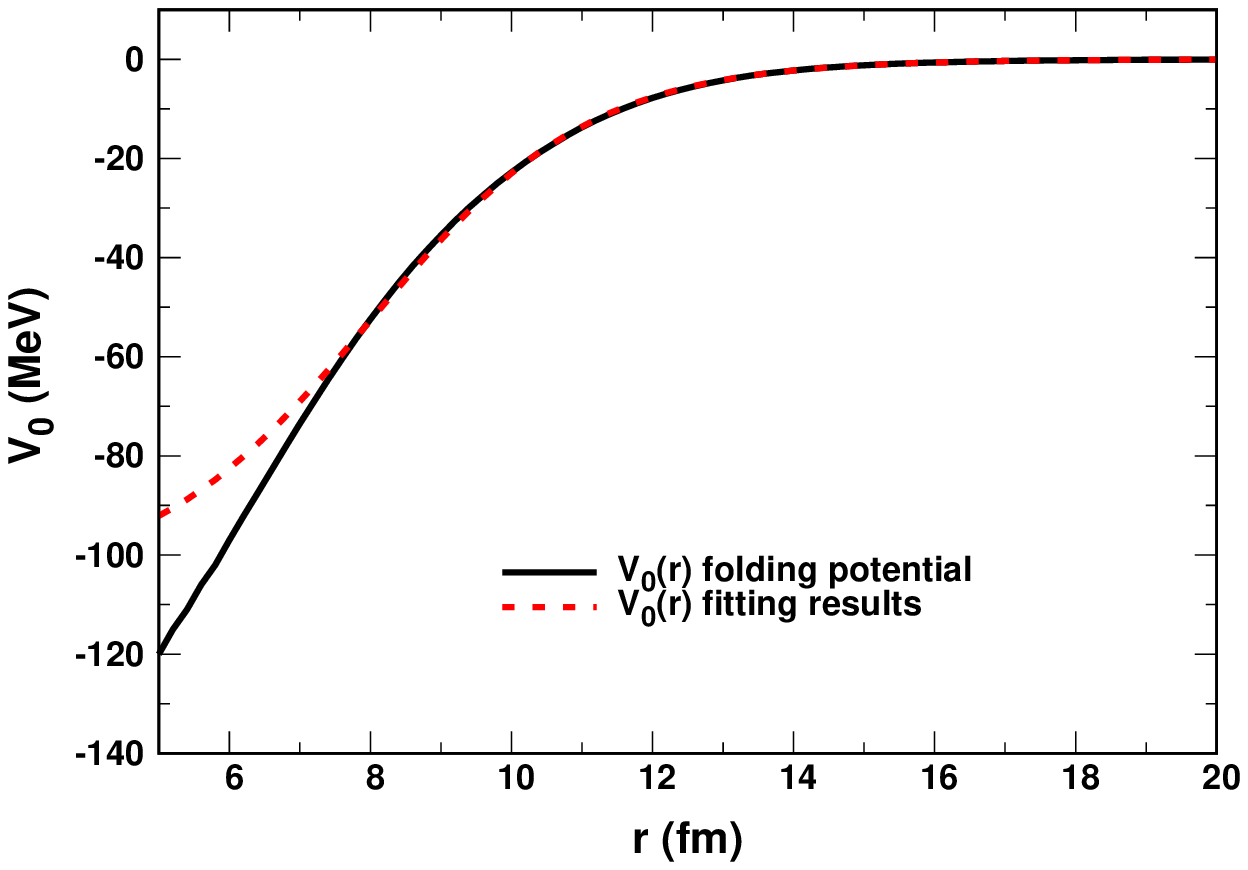} \\ \includegraphics[width=0.70\linewidth]{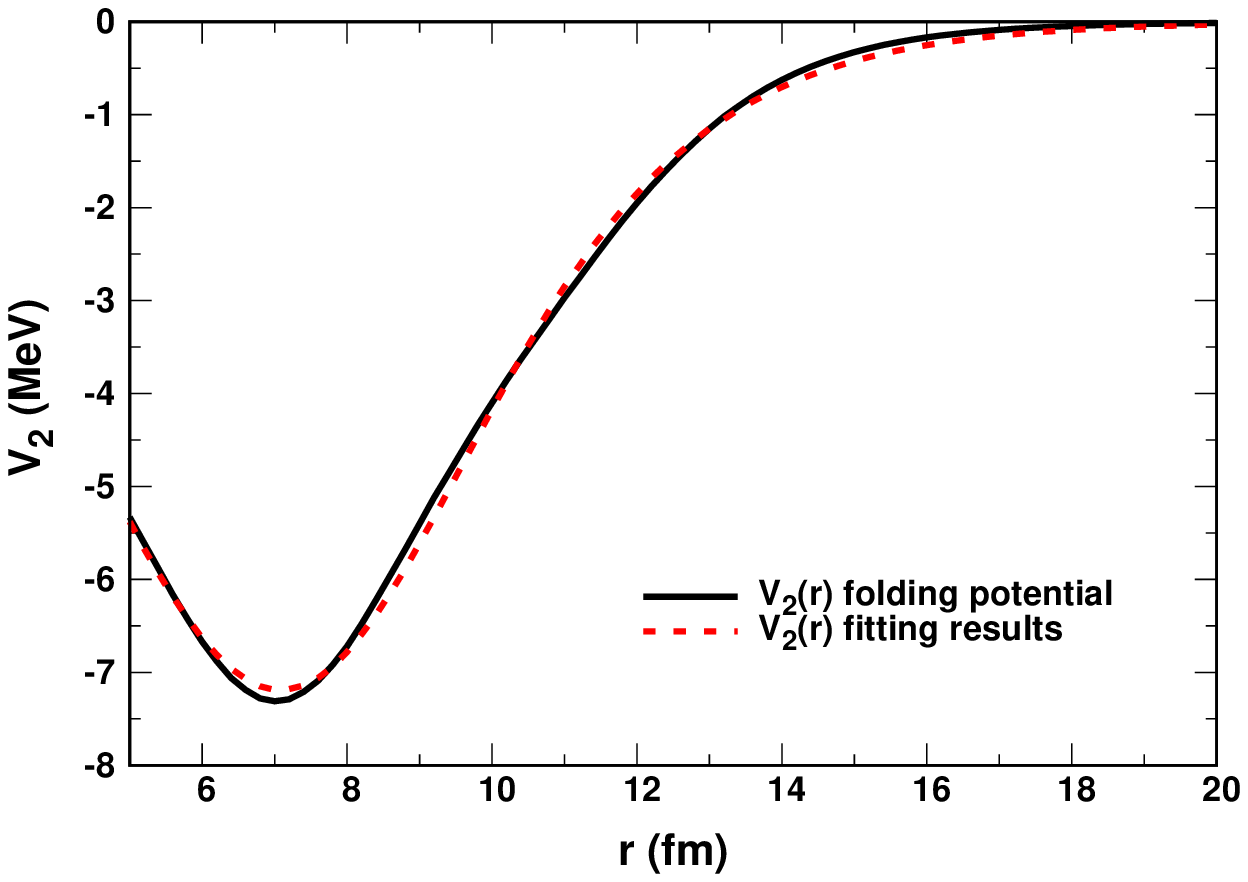} \\
\includegraphics[width=0.70\linewidth]{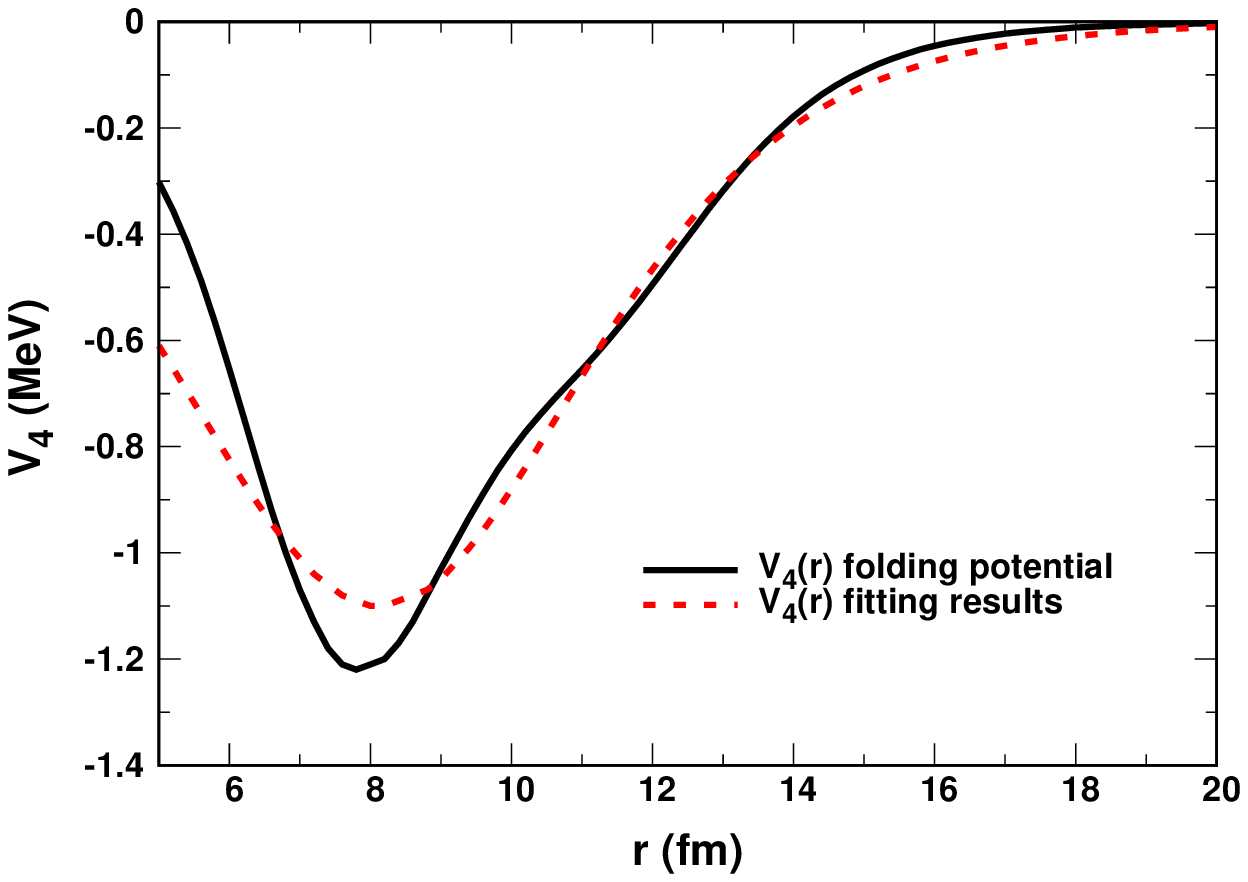}
\end{tabular}
\caption{The potential between the valence neutron in $^{15}$C
and the target nucleus $^{232}$Th for the  $^{15}$C + $^{232}$Th reaction.
The top, the middle, and the bottom panels are for
$V_{0}(r)$, $V_{2}(r)$, and $V_{4}(r)$ in Eq.~(\ref{expansion}), respectively.
The black solid lines  show the results of the double folding potential,
while the red dashed lines shows fits with the Woods-Saxon function.
}
\label{fig:3}
\end{figure}

The solid lines in Fig.~\ref{fig:3} shows the neutron-target potential obtained with the double
folding procedure. The top, the middle, and the bottom panels show the monopole,
the quadrupole, and the hexadecapole components, respectively.
In order to discuss properties of these potentials, we fit them with a Woods-Saxon
function and its first derivative.
That is,
\begin{eqnarray}
V_\lambda(r)&=&
\frac{-V_0}{1+ \exp\left[ \left( r-R_0 \right)/a\right]}~~~(\lambda=0), \\
&=&
\frac{-V_0 \exp\left[ \left( r-R_0 \right)/a\right]}{\left(1+\exp\left[ \left( r-R_0 \right)/a\right]\right)^2}~~~(\lambda=2,4), \nonumber \\
\end{eqnarray}
The results of the fitting are shown in Fig.~\ref{fig:3} by the dashed lines
(see Table~\ref{parapameter-n-232Th} for the parameters).
Since the region around the position of the Coulomb barrier is most
important for fusion cross sections, the fitting are performed mainly in the
surface region, $r >$ 8 fm.
In Fig.~\ref{fig:3}, one can see that the folding potential can be well fitted with
the Woods-Saxon function.
The hexadecapole component, $V_{4}(r)$, has some deviation from the Woods-Saxon
function, but its contribution to the total potential is much smaller than
the monopole and the quadrupole components.
We also find that the contribution of $\lambda=6$, that is, $V_{6}(r)$, is negligible,
with a small depth size of about $-$0.1 MeV.
Note that the Coulomb barrier parameters, obtained by setting the deformation parameters $\beta_{T\lambda}$ to be zero
in the folding procedure, are $V_{b}$=58.81 MeV, $R_{b}$=12.33 fm, and the $\hbar \Omega$=4.10 MeV,
which can be compared to those
for the $^{14}$C+$^{232}$Th system
listed in Table~\ref{set-14C}.
The Coulomb barrier height is lowered by 1.53 MeV
owing to the weakly bound valence neutron in $^{15}$C.

\begin{table}
\begin{ruledtabular}
\caption{Parameters for the neutron-target part of the
nuclear potential for the $^{15}$C+$^{232}$Th system.
Those are obtained by fitting the double folding potential to
the Woods-Saxon function and its derivative.
The depth, $V_{0}$, the radius, $R_{0}$, and the diffuseness, $a$, parameters
are shown for each multipole component. }
\label{parapameter-n-232Th}
\begin{tabular}{cccc}
   $V_\lambda$         &$V_{0}$ (MeV)   &  $R_{0}$ (fm)  &  $a$ (fm) \\ \hline
   $V_{0}(r)$  &     105.923   &    7.981     &  1.572       \\
   $V_{2}(r)$  &     28.756    &    7.080     &  1.890        \\
   $V_{4}(r)$  &      4.398    &    8.131    &   1.944        \\
\end{tabular}
\label{table2}
\end{ruledtabular}
\end{table}

\begin{figure}[h]

\includegraphics[width=0.70\linewidth]{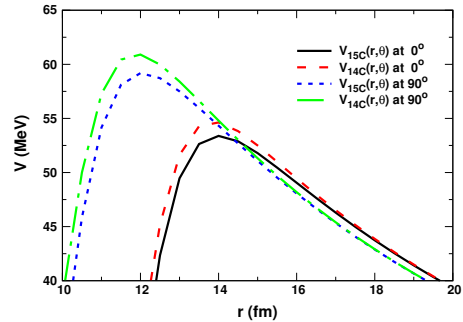}

\caption{The total potentials (the sum of the nuclear
and the Coulomb potentials) for the $^{14}$C + $^{232}$Th and the
$^{15}$C + $^{232}$Th systems for
$\theta = 0^o$ and  $\theta = 90^o$.
The solid and the dotted lines are for the $^{15}$C + $^{232}$Th system,
for which the solid line corresponds to $\theta = 0^o$ and the
dashed line corresponds to $\theta = 90^o$.
The dashed and the dot-dashed lines are the same as the solid and the
dotted lines, but for the $^{14}$C + $^{232}$Th system.
}
\label{fig:4}
\end{figure}

Figure~\ref{fig:4} shows the total potential (that is, the sum of the nuclear
and the Coulomb potentials).
The red dashed and the green dot-dashed lines show the potential for the
$^{14}$C+$^{232}$Th reaction for $\theta = 0^o$  and at $\theta = 90^o$,
respectively.
Note that the case of $\theta=0^o$ is referred to as a tip collision while
that of $\theta=90^o$ is referred to as a side collision.
The black solid and the blue dotted lines show the corresponding
potentials for the $^{15}$C+$^{232}$Th reaction, obtained with the Woods-Saxon fit to the double folding
potential between the valence neutron and the core nucleus.
For both the cases,
the Coulomb barrier is significantly lowered
due to the addition of the valence neutron in $^{15}$C.
It is well known that lowering of the Coulomb barrier leads to an increase of the penetration probability for fusion cross sections.

\subsection{Fusion cross sections}

Let us now calculate fusion cross sections for the $^{15}$C + $^{232}$Th system
using the potential constructed in the previous subsection.
For simplicity, we include up to $\lambda=4$ in the multipole expansion
of the double folding potential between the valence neutron in $^{15}$C and
the target nucleus.

\begin{figure}[h]
 \includegraphics[width=0.7\linewidth]{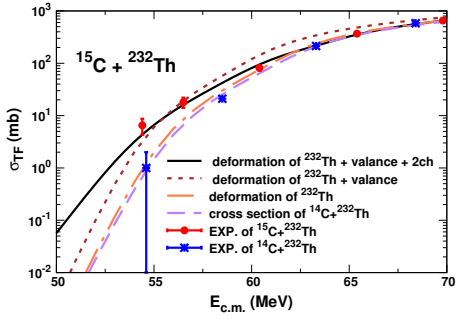}
 \caption{Fusion cross sections for the $^{15}$C + $^{232}$Th system.
The dot-dashed line shows fusion cross sections obtained by scaling
the potential for the  $^{14}$C + $^{232}$Th system with the mass number
of the colliding nuclei.
The dotted line shows fusion cross sections with the double folding potential,
which takes into account the halo structure of the $^{15}$C nucleus, while
the solid line shows
the result of the coupled-channels calculation
with the transfer coupling.
For comparison, the figure also shows the
theoretical fusion cross sections for the $^{14}$C + $^{232}$Th system
by the dashed line.
The experimental data are taken from Ref.~\cite{PhysRevLett.106.172701} .}
\label{fig:5}
\end{figure}

Figure~\ref{fig:5} presents a comparison between the calculated fusion cross sections
and the experimental data.
The red circles and the blue stars show the experimental data
for the $^{15}$C + $^{232}$Th and $^{14}$C + $^{232}$Th systems, respectively.
The violet dashed curve shows the results for the $^{14}$C + $^{232}$Th system,
which is the same as the solid line in Fig.\ref{fig:1}.
The coral dot-dashed curve
shows the results for the $^{15}$C + $^{232}$Th system, obtained using the potential
which is simply scaled from that for the $^{14}$C + $^{232}$Th system.
That is, the radius parameter in the Woods-Saxon potential for the
 $^{14}$C + $^{232}$Th system is changed
from $r_{0}\left(14^{1/3} + 232^{1/3}\right)$ to $r_{0}\left(15^{1/3} + 232^{1/3}\right)$.
This calculation does not take into account the weakly bound nature of the $^{15}$C
projectile, and corresponds to the calculation presented in Ref.
~\cite{PhysRevLett.106.172701}.
In fact, the dashed and the dot-dashed lines show
similar fusion cross sections to each other, as has been argued in
Ref. ~\cite{PhysRevLett.106.172701}. Even though this calculation
reproduces the experimental
data at energies above the Coulomb barrier, $E_{\rm c.m.}$ $\geqq$ 60 MeV,
it considerably underestimates fusion cross sections in the energy region
below the Coulomb barrier.

Fusion cross sections evaluated with the halo nature of the $^{15}$C nucleus
are shown by the brown dotted line.
This result clearly shows an enhancement of fusion cross sections with respect
to the dot-dashed line, and reproduce the experimental data at the
two lowest energies.
However, the fusion cross sections
in the region of 60 MeV $\lesssim E_{\rm c.m.} \lesssim$ 65 MeV
are clearly overestimated. 

In order to investigate a possible origin for the discrepancy, we
follow Ref. \cite{CHOI2018455} and consider a transfer coupling
to the $^{14}$C+$^{233}$Th channel.
That is, we consider a transfer to
a single effective channel \cite{Rowley2001}
and solve the coupled-channels equations
\cite{HT12,HRK99} of

\begin{eqnarray}
&&\left(\begin{array}{cc}
K+V_1(r, \theta) & F_{  1 \rightarrow 2 }(r)  \\
F_{ 1 \rightarrow 2 }(r) & K+V_2(r, \theta)-Q  \\
\end{array}\right)
\left(\begin{array}{c}
\psi_1(r)  \\
\psi_2(r)  \\
\end{array}\right)
\nonumber \\
&&\hspace*{4.5cm}=E
\left(\begin{array}{c}
\psi_1(r)  \\
\psi_2(r)  \\
\end{array}\right).
\label{eq:16}
\end {eqnarray}
Here, the channels 1 and 2 denote the $^{15}$C+$^{232}$Th
and the $^{14}$C+$^{233}$Th systems, respectively.
$K$ is the kinetic energy (with the centrifugal potential)
and $V_i(r,\theta)$ ($i$=1,2) is the inter-nucleus potential for each
partition.
$Q$ is the effective $Q$-value for the one-neutron transfer process,
while
$F_{ 1 \rightarrow 2 }(r)$ is the coupling form factor.
Notice that the effective transfer channel may mock up also the breakup
channel to some extent.

\begin{table}[h]
\begin{ruledtabular}
\begin{center}
\caption{The parameters for the transfer coupling.
Here, those in the coupling form factor, Eq. (\ref{eq:17}), are determined by fitting
the results of the coupled-channels calculations to the experimental
fusion cross section for the $^{15}$C+$^{232}$Th system
with $Q=0$ for the transfer $Q$-value.  
}
\begin{tabular}{cccc}
$Q$ (MeV) & $F_{t}$ (MeV fm) & $R_{\rm coup}$ (fm) &   $a_{\rm coup}$ (fm)
\\ \hline
0 & 27.5  & 14.638 & 0.69               \\
\end{tabular}
\end{center}
\end{ruledtabular}
\end{table}

In the calculation,
we assume that the potential $V_2(r, \theta)$ for the $^{14}$C + $^{233}$Th
channel is the same as the potential for the
$^{14}$C + $^{232}$Th  presented in Table \ref{set-14C}.
For the coupling form factor,
$F_{ 1 \rightarrow 2 }(r)$,
we employ the derivative form of
the Woods-Saxon potential~\cite{DV86,dasso1985macroscopic} given by
\begin{equation}
F_{1\rightarrow 2}(r)=F_t \,
\frac{d}{dr}\left( \frac{1}{1+\exp((r-R_{\rm coup})/a_{\rm coup})} \right).
\label{eq:17}
\end {equation}
The parameters are determined by fitting to the experimental
data for fusion cross sections (see Table III).
To this end, we take the transfer $Q$-value to be zero, $Q=0$,
rather than the ground-state-to-ground-state $Q$-value, $Q_{\rm gg}=+3.568$ MeV,
taking into account the $Q$-value matching condition  \cite{Rowley2001}.
The black solid line in Fig. \ref{fig:5}
shows the fusion cross sections so obtained.
One can see that the experimental data are well reproduced in the whole
energy region shown in the figure, indicating an importance of
the dynamical effect on fusion of the neutron-rich nucleus, $^{15}$C.


\section{Summary}
\label{sec4}

We have calculated fusion cross sections for the $^{15}$C+$^{232}$Th system,
for which $^{15}$C is a well-known one-neutron halo nucleus
while $^{232}$Th is a well deformed nucleus.
To this end,
we have evaluated the cross sections within the double folding formalism
taking into account the halo structure of $^{15}$C and the deformation of the
target nucleus.
In addition, we have also taken into account the coupling to the one-neutron
transfer channel to the $^{14}$C+$^{233}$Th system with the coupled-channels
formalism.
We have shown that such calculation reproduces simultaneously
well the experimental data for the $^{15}$C+$^{232}$Th
and $^{14}$C+$^{232}$Th systems.
This clearly indicates that
all of the halo structure of $^{15}$C, the deformation of $^{232}$Th, and
the dynamical effect such as transfer and breakup play an important role in
fusion of the $^{15}$C+$^{232}$Th system.
In this regard, it would be useful to investigate systematically fusion
of a halo nucleus with deformed target nuclei in order to
gain a deeper insight into the role of deformation of the target
in fusion of neutron-rich nuclei. 
A comparison
of the $^{15}$C + $^{232}$Th system
to the
$^{15}$C + $^{238}$U and  $^{15}$C+$^{208}$Pb systems
might also provide useful information.

\section*{Acknowledgment}

This work was supported by the National Research Foundation of Korea (Grant No. NRF-2016R1C1B1012874, 2018R1D1A1B07045915, NRF-2017R1E1A1A01074023,NRF-2019R1H1A1102164, NRF-2020R1A2C3006177 and NRF-2013M7A1A1075764) 
and by JSPS KAKENHI (Grant No. JP19K03861). 
K. S. Kim's work was supported by MSIT (No.2018R1A5A1025563).

\end{document}